\documentstyle[12pt]{article}

\def\[{\left[}
\def\]{\right]}

\def\a{\begin{eqnarray}}
\def\b{\end{eqnarray}}
\def\0{\nonumber}
\input amssym.def
\newfam\eusmfam
\def\sG{sine--Gordon}
\def\Lio{Liouville}
\def\pde{partial differential equation}
\def\pdes{partial differential equations}
\def\Tsc{Tchebyscheff}
\renewcommand{\theequation}{\thesection.\arabic{equation}}

\newlength{\extraspace}
\newlength{\extraspaces}
\newcounter{dummy}

\newcommand{\ai}{
\addtocounter{equation}{1}
\setcounter{dummy}{\value{equation}}
\setcounter{equation}{0}
\renewcommand{\theequation}{\thesection.\arabic{dummy}\alph{equation}}
\begin{eqnarray}
\addtolength{\abovedisplayskip}{\extraspaces}
\addtolength{\belowdisplayskip}{\extraspaces}
\addtolength{\abovedisplayshortskip}{\extraspace}
\addtolength{\belowdisplayshortskip}{\extraspace}}
\newcommand{\bj}{
\end{eqnarray}
\setcounter{equation}{\value{dummy}}
\renewcommand{\theequation}{\thesection.\arabic{equation}}}

\input amssym.tex
\newcommand{\be}{\begin{equation}}
\newcommand{\ee}{\end{equation}}
\newcommand{\ba}{\begin{eqnarray}}
\newcommand{\ea}{\end{eqnarray}}
\newcommand{\ban}{\begin{eqnarray*}}
\newcommand{\ean}{\end{eqnarray*}}
\newcommand{\brr}{\begin{array}}
\newcommand{\err}{\end{array}}
\newcommand{\bc}{\begin{center}}
\newcommand{\ec}{\end{center}}

\def\D{{\cal D}}
\def\E{{\cal E}}

\def\G{{\cal G}}

\def\S{{\cal S}}

\def\l{\lambda}
\def\s{\sigma}
\def\al{\alpha}
\def\be{\beta}
\def\ga{\gamma}
\def\de{\delta}
\def\ep{\epsilon}
\def\o{\omega}

\def\Ga{\Gamma}
\def\t{\theta}
\def\var{\varphi}

\def\var{\varphi}
\newcommand{\bea}{\begin{eqnarray}}
\newcommand{\eea}{\end{eqnarray}}
\newcommand{\bean}{\begin{eqnarray*}}
\newcommand{\eean}{\end{eqnarray*}}
\newcommand{\RR}{\Bbb R}
\newcommand{\CC}{\Bbb C}

\newcommand{\DD}{\Bbb D}
\newcommand{\HH}{\Bbb H}
\newcommand{\LL}{\Bbb L}
\newcommand{\PP}{\Bbb P}

\newcommand{\FF}{\Bbb F}

\newcommand{\del}{\partial}

\begin{document}

\begin{titlepage}
\begin{flushright}
IFT-P.070/99\\
CBPF--NF--045/99
\end{flushright}

\vskip0.5cm
\centerline{\LARGE Surfaces of Constant negative Scalar Curvature }

\centerline{\LARGE and the  Correpondence between the \Lio\,}

\centerline{\LARGE and the \sG\, Equations}
\vskip1.5cm
\centerline{\large   H. Belich$^a$, G. Cuba$^a$} 
\vskip0.5cm
\centerline{\large and }
\vskip0.5cm
\centerline{\large R. Paunov$^{a,b}$
\footnote{Corresponding author E--mail address paunov@cbpfsu1.cat.cbpf.br}}
\centerline{$^a$Centro Brasileiro de Pesquisas Fisicas }
\centerline{Rua Dr. Xavier Sigaud 150, Rio de Janeiro, 22290--180, RJ Brazil}
\vskip0.5cm
\centerline{$^b$ Instituto de F\'{\i}sica Te\'orica--IFT/UNESP}
\centerline{Rua Pamplona 145, 01405--900 S\~ao Paulo, SP, Brazil}

\vskip2.5cm
\abstract{By studying the {\it internal} Riemannian geometry of the surfaces of constant negative scalar curvature, we obtain a natural map between the \Lio\, and the sine--Gordon equations. First, considering isometric immersions into the Lobachevskian plane, we obtain an uniform expression for the general (locally defined) solution of both the equations. Second, we prove that
there is a Lie--B\"acklund transformation interpolating between Liouville and 
sine--Gordon. Third, we use isometric immersions into the Lobachevskian plane to describe sine--Gordon $N$--solitons explicitly.}

\vskip1.5cm
PACS numbers: 03.65. Pm, 11.10 Lm, 02.30 Jr, 02.20 Tw

\end{titlepage}
\clearpage
\pagestyle{myheadings}
\markright{\hskip1.5cm H. Belich, G. Cuba and  R. Paunov,
Surfaces of Constant Negative ...}

\renewcommand{\theequation}{\thesection.\arabic{equation}}
\setcounter{equation}{0}
\setcounter{subsection}{0}
\setcounter{footnote}{0} 
\section{Introduction}

The problem of searching the geometrical origin of a given physical theory is of great interest, because its solution gives a better understanding, and perhaps 
an exact solution of the underlying model. In the present paper we study the geometry of two physically relevant \pdes\,: the \Lio\, and the \sG\, ones.
Both the equations has been obtained in he previous century by studying the 
differential geometry of surfaces in $\RR^3$ \cite{Ei,Dub}. The geometrical meaning of these equations is similar: the \Lio\ equation describes minimal surfaces, i.e. surfaces with vanishing mean curvature, while the \sG\, equation is related to surfaces of constant negative scalar curvature. The \Lio\, equation appears also in the uniformization theory of the Riemann surfaces.
Many--valued surface transformations between surfaces of constant scalar negative curvature 
has been studied by Bianchi, Lie, B\"acklund and Darboux \cite{Bi} (for a more recent review on the subject, see \cite{An, Ab,Ten}). The transformation originally discussed by B\"acklund in 1880 which maps the \sG\, into itself, is a particular example of what nowadays is called a B\"acklund transformation. The latter play important role in the theory of the \pdes\, solvable by Inverse Scattering Method \cite{Ab}.

In  physical applications, the interest to the quantum \Lio\, model is motivated by its relation to the string theory, conformal field theory in two dimensions 
and 2$D$ gravity (for a review see \cite{Pol}). The classical equations of motion of a string in a flat target space describe a surface with vanishing mean curvatures. In contrast to the classical equations of motion, within the quantum theory, the \Lio\, action arises as a (Weyl) anomaly of Polyakov's path integral after integration over the world sheet metric in the conformal gauge. 

On the other hand, both \Lio\, and \sG\, are examples of completely integrable (in the sense of \Lio\,)  field theories. In particular, the field equations admit a zero curvature representation. An important property of the \sG\, equation is that it has {\it soliton} solutions \cite{Ab,inv} which describe elastic collision of localizable waves. The latter can be interpreted as new particles of nonperturbative nature which appear in the spectrum of the theory. 
Due to the quantum integrability, the scattering between the 
quantum \sG\, solitons remains elastic. Exact $S$ matrix which factorizes into two
soliton interactions has been proposed in \cite{Zam}.

The idea to construct integrable models in two dimensions by studying embeddings of surfaces in varieties of higher dimensions has a huge history. In its whole generality it was formulated by Saveliev \cite{Sav} who considered embeddings in spaces with a fixed (simple) group of motions. The proposal advanced in \cite{Sav} is to consider the equations of Gauss, Codazzi and Ricci \cite{Gauss,Will} which describe embeddings of some fixed    two dimensional manifold into a space of higher dimension and  to select "integrable" embeddings. By integrability one understands that  the embedding equations are equivalent to the zero curvature condition of a certain (Lax) connection. The classification of the integrable embeddings is a rather involved problem since  the equations of Gauss, Codazzi and Ricci, in general  are very complicated, and thus,  
difficult to solve. Some special embeddings into the three dimensional affine space has been considered in \cite{Ga} in relation to the $W_3$ generalization
of the Polyakov's gravity \cite{Pol}. The canonical  Lax pair of the $A_n$ 
Toda theories was derived by studying the extrinsic geometry of surfaces which are "chirally" embedded in $\CC\PP^n$ \cite{Ger}. Surfaces embedded in $\RR^3$
which admit nontrivial ismetries preserving the mean curvature has been 
recently considered in \cite{Bob} in relation to the Bonnet problem.

In the present paper we consider the {\it internal} Riemannian geometry of surfaces of constant negative scalar curvature $R=-2$. It is well known that fixing conformal coordinates on the surface, one derives the \Lio\, equation (\ref{2.9a}), whereas in the generalized \Tsc\, coordinates, the \sG\, (\ref{2.9b}) equation appears. This simple observation yields to the conclusion that there should be  a
(locally) invertible transformation which maps the \Lio\, equation into the \sG\, one. The latter relation can be in principle derived by solving the Laplace--Beltrami equation associated to the \Tsc\, metric (\ref{2.8a}). To get explicit expressions, we consider isometric immersions in the Lobachevskian plane. There is a standard theorem \cite{Dub} which guarantees that such
immersion always exits locally and it is fixed up an isometry transformation of the Lobachevskian plane. These immersions produce naturally solutions of the 
\Lio\, and \sG\, equations. In conformal coordinates, we recover the famous \Lio\, formula ( see for example \cite{FT,L}). Using the zero curvature representation, we show that the above mentioned isometric immersions are expressed in terms of the entries of special matrix solutions of the underlying linear problems. We further study the isometry maps in the Lobachevskian plane by fixing conformal and \Tsc\, coordinates on a given surface of constant negative scalar curvature. This gives us a field dependent and nonlocal (that is, depending on the derivatives of arbitrary order) change of the local coordinates which produces a Lie--B\"acklund transformation between the \Lio\, and the \sG\, equation.
The transformation between these two integrable models has been previously discussed in the literature \cite{Dol} in relation to the instanton solutions of $4$--$D$  $SU(2)$ Yang--Mils equations with cylindrical symmetry \cite{Witt}. Finally, we study the image of the \sG\, solitons in the Lobachevskian plane.

This paper is organized as follows. In section 2 we review some basic facts concerning the geometry of the surfaces of constant negative scalar curvature. Solutions of \Lio\, and \sG\, equations are obtained by local isometric immersions in the Lobachevskian plane. In section 3, by using the zero curvature representation, we show that these are the general solutions of both the equations. In section 4 we construct a Lie--B\"acklund transformation which interpolates between \Lio\, and \sG . In section 5 we study isometric immersions in the Lobachevskian plane which correspond to $N$--soliton solution of the \sG\, equation.
 
\renewcommand{\theequation}{\thesection.\arabic{equation}}
\setcounter{equation}{0}
\setcounter{subsection}{0}
\setcounter{footnote}{0} 
\section{Geometric Origin of \Lio\, and \sG\,}

The aim of this section is to review the geometric interpretation of the \Lio\,
and  the \sG\, equations. It is well known that both the equations appear in studying the Riemannian geometry of surfaces of constant negative scalar curvature. The latter are also known as {\it pseudospherical surfaces}.
Within  the classical differential geometry, surfaces of constant negative scalar curvature 
are usually considered as varieties embedded (with the induced natural Riemannian metric) into the three dimensional Euclidean space $\RR^3$. The 
underlying Riemannian structure on the surface is determined by the equations 
of Gauss, Codazzi and Ricci \cite{Ten,Will}. Here, in contrast to the classical treatment, we shall focus our attention on the internal Riemannian geometry of the pseudospherical surfaces. The latter admit (locally) an isometric immersion
in the Lobachevskian plain $\HH$. The study of these immersions, allows  to construct explicit solutions  of the 
\Lio\, and the \sG\, equations. 

We  recall some basic definitions and notions of the differential geometry of surfaces \cite{Ei,Dub,Ten,Will}.
Let $\S$ be a two dimensional smooth manifold. We fix local coordinates $x^i$,
$i=1,2$ on $\S$ and denote by $\del_i=\frac{\del}{\del x^i}$ the tangent vectors related to the corresponding coordinate frame. The $1$--forms $dx^i$ form a basis of the cotangent space which is dual to $\{\del_i\}$: $dx^i(\del_j)=\delta^i_j$. Any vector field $X$ and any $1$--form $\al$ can be written as $X=X^i\del_i$ and $\al=\al_idx^i$ respectively. A Riemannian structure on $\S$ is induced by a symmetric positive definite metric $ds^2=g_{ij}dx^idx^j$, $g_{ij}=g_{ji}$. Here we perform a slight abuse of terminology, since by definition, the Riemannian structure, is 
introduced as a class of isometric Riemann manifolds. The metric
on $\S$ allows to introduce a symmetric inner product on the tangent bundle $T\S$: 
$<X,Y>=g_{ij}X^iY^j$. Let $\nabla : T\S\times T\S \rightarrow T\S$ be an
affine  connection \cite{Dub,Will} on  $T\S$. The curvature 
and the torsion are given by the standard expressions
\ai
&&R(X,Y)Z=\nabla_X\nabla_YZ-\nabla_Y\nabla_XZ-\nabla_{\[X,Y\]}Z\0\\
&&T(X,Y)=\nabla_XY-\nabla_YX-\[X,Y\]
\label{2.1a}
\b
where $X,\, Y,\, Z$ are vector fields and
\a
&&\[X,Y\]=\left(X^j\del_jY^i-Y^j\del_jX^i\right)\del_i\0
\b
is the Lie bracket between  $X$ and $Y$. In view of the standard
properties of the affine connection $\nabla$, it is not difficult to establish 
the tensorial nature of the curvature and the torsion. The components of the curvature and the torsion tensor are given by
\a
&&R_{ijkl}=<R(\del_k , \del_l)\del_j , \del_i>\0\\
&&T^k_{ij}=dx^k\left(T(\del_i,\del_j)\right)
\label{2.1b}
\bj
respectively. The scalar curvature
\a
&&R=g^{ik}g^{jl}R_{ijkl}
\label{2.2}
\b
where $g^{ij}$ is the inverse of the metric tensor $g^{ij}g_{jl}=\delta^{i}_{l}$, is invariant under changes of the local coordinates.

A Riemannian manifold of arbitrary dimension admits an unique torsionless
connection
\ai
&&T(X,Y)=0\label{2.3a}
\b
such that
\a
X<Y,Z>=<\nabla_X Y,Z>+<Y,\nabla_X Z>
\label{2.3b}
\bj
for arbitrary vector fields $X$, $Y$ and $Z$ on $\S$. This connection is known in the literature \cite{Dub,Will} as the {\it Levi--Civita} connection. 

Now we are in a position to study surfaces of constant negative scalar curvature. In the present 
paper we will be interested on Riemann structures induced by a metric 
of the following form
\a
&&ds^2=g_{11}(dx^1)^2+2g_{12}dx^1dx^2+g_{22}(dx^2)^2\0\\
&&\del_i g_{jj}=0,\quad i,j=1,2
\label{2.4}
\b
where the local coordinates $x^i$ are not 
necessarily real. In view of (\ref{2.3a}), one gets the identity
\ai
&&\nabla_1\del_2=\nabla_2\del_1,\quad
\nabla_i=\nabla_{\del_i}\label{2.5a}
\b
On the other hand, the diagonal components of the metric (\ref{2.4}) 
$g_{ii}=<\del_i , \del_i>$ are constants. Thus, taking into account (\ref{2.3b}), one concludes that
\a
\nabla_i \del_j=\nabla_j\del_i=0\quad i\neq j
\label{2.5b}
\b
The above identities admit a clear geometrical interpretation: the coordinate vector field $\del_2$ ($\del_1$) is {\it parallel transported} along the 
vector field $\del_1$ ($\del_2$) with respect to the Levi--Civita connection 
$\nabla$. Taking into account (\ref{2.3b}), (\ref{2.4}) and the above identities,
it is  not difficult to get the expressions \cite{Dub}
\a
&&\nabla_1\del_1=\frac{1}{g}\left( -g_{12}\del_1 g_{12}\,\del_1+g_{11}\del_1g_{12}\,\del_2\right)\0\\
&&\nabla_2\del_2=\frac{1}{g}\left(g_{22}\del_2 g_{12}\,\del_1-
g_{12}\del_2g_{12}\,\del_2\right)\0\\
&&g={\rm det}\left(g_{ij}\right)= g_{11}g_{22}-g_{12}^2
\label{2.5c}
\bj
Due to (\ref{2.1a}), (\ref{2.1b}), (\ref{2.5b}) and the 
above identities, one gets
\ai
&&R_{1212}=\del_1\del_2 g_{12}+\frac{g_{12}}{g}\del_1g_{12}\del_2g_{12}
\label{2.6a}
\b
We recall the symmetries of the Riemann
tensor (\ref{2.1b}) associated to the Levi--Civita connection: $R_{ijkl}=-R_{jikl}=-R_{ijlk}$, $R_{ijkl}=R_{klij}$ \cite{Dub,Will}.
Therefore, in two dimensions, the Riemann tensor has only one independent component: $R_{1212}$.
In particular, the scalar curvature ({\ref{2.2})  can be written as
\a
&&R=\frac{2}{g}R_{1212}
\label{2.6b}
\bj

Let us first fix conformal frame on the surface $\S$.
Setting $x^1=z$, $x^2=\bar z$ where $z$ and $\bar z$ are complex coordinates, the metric is the following
\ai
&&ds^2=e^{\var(z,\bar z)}dzd\bar z
\label{2.7a}
\b
It is a well known fact in the theory of surfaces \cite{Ei,Dub} that any 
Riemannian metric on $\S$ is conformally flat, i. e. by a suitable change of 
the local coordinates, it reduces to the above expression.
Since a conformally flat metric is a particular case of (\ref{2.4}),
one can use (\ref{2.6a}) to calculate the scalar curvature (\ref{2.6b}).
The result is \cite{Dub}
\a
&&R=-4e^{-\var(z,\bar z)}\del\bar \del \var(z,\bar z)\0\\
&&\del=\frac{\del}{\del z},\quad \bar\del=\frac{\del}{\del \bar z}
\label{2.7b}
\bj

Another possible choice is to consider {\it generalized} \Tsc\, coordinates on $\S$
\ai
&&ds^2=\l^2(dx^+)^2+\l^{-2}(dx^-)^2+ 2\cos\psi dx^+ dx^-
\label{2.8a}
\b
where $x^{\pm}$ are real local coordinates, $\psi=\psi(x^+,x^-)$ is a real function and $\l$ is real constant. Inserting again the general expressions
(\ref{2.6a}) and (\ref{2.6b}) into (\ref{2.8a}) one concludes that the scalar curvature of the generalized \Tsc\, metric is
\a
&&R=-2\frac{\del_+\del_-\psi}{\sin \psi},\quad
\del_{\pm}=\frac{\del}{\del x^{\pm}}\label{2.8b}
\bj
Imposing the condition that $\S$ is a surface of constant negative scalar curvature $R=-2$, one deduces from (\ref{2.7b}) the \Lio\, equation
\ai
&&\del\bar \del \var=\frac{1}{2}e^{\var}\label{2.9a}
\b
whereas (\ref{2.8b}) yields the \sG\, equation
\a
&&\del_+\del_-\psi=\sin \psi \label{2.9b}
\bj
Therefore the two above equations admit a clear geometrical interpretation.
More precisely, they appear by fixing special local coordinate frames on pseudospherical surfaces.

At this stage the question of existence of a "privileged" surface $\HH$ of scalar
curvature $R=-2$ arises. By "privileged" we understand that any other surface
$\S$ of the same scalar curvature admits, at least locally, an isometric immersion 
$\S\rightarrow^{\hskip -.3cm i}\,\,\HH$. In particular, the metric on $\S$ is a pull-back of the metric on $\HH$. The answer of the above question is positive
\cite{Dub}: as $\HH$ one can choose the Lobachevskian plane
$\HH=\{u\in\CC\,|{\rm Im}u>0\}$ equipped with the metric
\ai
&&ds^2=-4\frac{du\,d\bar u}{(u-\bar u)^2}
\label{2.10a}
\b
In view of (\ref{2.7b}), $\HH$ is a variety of constant negative scalar curvature $R=-2$. Moreover the expression   
\a
&&e^{\var(u,\bar u)}=-\frac{4}{(u-\bar u)^2}
\label{2.10b}
\bj
satisfies the \Lio\, equation (\ref{2.9a}) with respect to the complex variables $u$ and $\bar u$ (${\rm Im} u>0$). Suppose now that $\S$ is a pseudospherical
surface and $(z,\bar z)$ are conformal coordinates on it. From (\ref{2.10a}) it is seen that $u$ has to be holomorphic or antiholomorphic function on $z$. Since
the Lobachevskian metric is invariant under the exchange $u\leftrightarrow 
\bar u$, we shall assume in what follows that $u=u(z)$, $\bar u= \bar u (\bar z)$. Therefore,  from (\ref{2.10a}) it follows that the induced metric on $\S$ is given by
\a
&&ds^2=e^{\var(z,\bar z)}dz d\bar z,\0\\
&& e^{\var(z,\bar z)}=-4\frac{\del u\, \bar\del \bar u}{(u-\bar u)^2},
\quad \bar \del u=\del \bar u =0
\label{2.11}
\b
The above expression for the \Lio\, field is the famous \Lio\, formula
\cite{FT}. It implies in particular that $e^{\var}$ is a $(1,1)$ form with 
respect to holomorphic (or conformal) changes of the local coordinates.
As a consequence, one recovers the conformal invariance of the \Lio\, equation
\a
&&z\rightarrow z^{\prime}=z^{\prime}(z),\quad
\bar z\rightarrow \bar z^{\prime}=\bar z^{\prime}(\bar z)\0\\
&&\var(z,\bar z) \rightarrow \var^{\prime}(z^{\prime}, \bar z^{\prime})=
\var(z,\bar z)+\ln \frac{ d z}{d z^{\prime}}+
\ln \frac{ d \bar z}{d \bar z^{\prime}}
\label{2.12}
\b

Let us now consider the \sG\, case. Suppose that the generalized \Tsc\, metric
(\ref{2.8a}) on  a surface $\S$ of constant negative scalar curvature is a pull-back of the Lobachevskian metric on $\HH$ (\ref{2.10a}). In particular, this wants to say that the map $\S\rightarrow^{\hskip -.3cm i}\,\,\HH$ satisfies the equations 
\a
&&e^{\pm i\psi}=-4\frac{\del_{\pm}u\, \del_{\mp}\bar u}{(u-\bar u)^2},\0\\
&&\hskip-1.5cm \del_{\pm}u\, \del_{\pm}\bar u=-\frac{\l^{\pm 2}}{4} (u-\bar u)^2,\quad\quad
u=u(x^+,x^-,\l), \quad \bar u = \bar u (x^+,x^-,\l) \label{2.13}
\b
which are the \sG\, counterpart of the \Lio\, formula (\ref{2.11}). It is instructive to check directly that the above expressions provide a solution of the \sG\, equation. To do that we first observe that the equations
\ai
&&\hskip-2.5cm \nabla_+\del_-=\nabla_-\del_+=
\left(\del_+\del_-u-2\frac{\del_+u\del_-u}{u-\bar u}\right)\frac{\del}{\del u}+
\left(\del_+\del_-\bar u+2\frac{\del_+\bar u\del_-\bar u}{u-\bar u}\right)
\frac{\del}{\del \bar u}
\label{2.14a}\\
&&\nabla_{\pm}\del_{\pm}=
\left(\del_{\pm}^2u-2\frac{(\del_{\pm}u)^2}{u-\bar u}\right)\frac{\del}{\del u}+
\left(\del_{\pm}^2\bar u+2\frac{(\del_{\pm}\bar u)^2}{u-\bar u}\right)\frac{\del}{\del u}\label{2.14b}
\bj 
are valid. In deriving these identities we have used (\ref{2.13}) as well 
as the covariant derivatives
\a
&&\nabla_u\frac{\del}{\del u}=-\frac{2}{u-\bar u} \frac{\del}{\del u}, \quad
\nabla_{\bar u}\frac{\del}{\del \bar u}=\frac{2}{u-\bar u} \frac{\del}{\del \bar u}\0\\
&&\nabla_u\frac{\del}{\del \bar u }=0, \quad 
\nabla_{\bar u}\frac{\del}{\del u }=0,
\label{2.15}
\b
which according to (\ref{2.5c}) define the Levi--Civita connection on the 
Lobachevskian plane. Due to (\ref{2.5b}), the covariant derivatives (\ref{2.14a}) vanish identically. Taking into account  this observation and and using (\ref{2.13}) we get
\a
&&i\del_+\psi=\frac{\del_+^2 u}{\del_+u}-2\frac{\del_+u}{u-\bar u}\0\\
&&i\del_-\psi=\frac{\del_-^2\bar u}{\del_-\bar u}+2\frac{\del_-\bar u}{u-\bar u}
\label{2.16}
\b
Using again (\ref{2.13}) and the fact that (\ref{2.14a}) are vanishing, we conclude that the above system is integrable and $\psi$ satisfies the \sG\, equation (\ref{2.9b}).

To close this section we shall make the following remark:
as it is seen from (\ref{2.11}) and (\ref{2.13}), an isometric immersion of 
a given pseudospherical surface ( with scalar curvature $R=-2$) yields to 
solutions of the \Lio\, and the \sG\, equations. On the other hand, an isometric immersion $\S\rightarrow^{\hskip -.3cm i}\,\,\HH$  is fixed up to an isometry transformation of $\HH$. It is well known that the group of the isometries of 
the metric (\ref{2.10a}) coincides with  $PSL(2,\RR)$. It acts on the upper half plane by projective (or M\"obius) transformations
\a
&&u\rightarrow \frac{\al u+\be}{\gamma u +\delta}\0\\
&&\al , \be , \gamma , \delta \in \RR, \quad \al \delta-\be\gamma =1
\label{2.17}
\b
A straightforward calculation shows that  the equations (\ref{2.11}) and (\ref{2.13}) are invariant with respect to the above transformation.

\renewcommand{\theequation}{\thesection.\arabic{equation}}
\setcounter{equation}{0}
\setcounter{subsection}{0}
\setcounter{footnote}{0} 
\section{General Solutions of \Lio\, and \sG\,}

The present section is devoted to the study of the general solutions of the \Lio\, (\ref{2.9a}) and of the \sG\, (\ref{2.9b}) equations. Our goal will be to show that the expressions  (\ref{2.11}) and (\ref{2.13})  exhaust, at least {\it locally} , the space of
solutions of (\ref{2.9a}) and (\ref{2.9b}) respectively. In view of the analysis
presented before, it turns out that any solution of the \Lio\, and the \sG\, equations can be described as an isometric immersion of a surface of constant negative scalar into the Lobachevskian plane $\HH$. Within this section we shall adopt 
a terminology borrowed from the string theory: $u$ and $\bar u$ (${\rm Im u}>0$) will be called "target space" variables; the local coordinates $(z,\bar z)$ which appear in the \Lio\, equation (\ref{2.9a}), as well as $x^{\pm}$ related to the \sG\, equation (\ref{2.9b}) will be referred to as "world--sheet" variables.

To show that (\ref{2.11}) and (\ref{2.13})
provide (at least locally)   general solutions of the corresponding \pdes , we
shall use the zero curvature representation. The \Lio\,
equation admits a zero curvature representation $F_{z\,\bar z}=\[\D_z\, ,\,
\D_{\bar z}\]=0$ for a connection which is in the Lie algebra $sl(2)$
\ai
&&\D_z=\del +A_z,\quad \D_{\bar z}=\bar{\del}+A_{\bar z}\0\\
&&\hskip-1.5cm A_z=\del \Phi +\frac{1}{2}e^{{\rm ad} \Phi}E^+,\quad
A_{\bar z}=-\bar{\del} \Phi + \frac{1}{2}e^{-{\rm ad} \Phi}E^-, \quad
\Phi=\frac{1}{4} \var H .\label{3.2a}
\b
In the above expressions $H$ and $E^{\pm}$ are the generators of $sl(2)$
\a
&&\[H, E^{\pm}\]=\pm 2 E^{\pm},\quad \[E^+, E^-\]=H .\0
\b
Similar representation is also valid for the \sG\, equation $F_{+-}=\[\D_+,\D_-\]=0$. The covariant derivatives $\D_{\pm}$ are given by
\a
&&\D_{\pm}=\del_{\pm}+A_{\pm},\quad 
A_{\pm}=\pm i \del_{\pm} \Psi +\frac{1}{2}e^{\pm  i {\rm ad}\Psi}\E_{\pm}\0\\
&&\Psi=\frac{1}{4}\psi H,\quad \E_{\pm}=\l^{\pm 1}(E^++E^-).\label{3.2b}
\bj
Due to the zero curvature condition, there exists a solution of the parallel transport equations
\ai
&&\D_{\al}\t =\left(\del_{\al}+A_{\al}\right)\t=0\label{3.3a}
\b
where $\al=z, \bar z$ for (\ref{3.2a}) and $\al=\pm$ for (\ref{3.2b}).
Within the Inverse Scattering Method \cite{Ab,inv}, the above equation is known as the auxiliary linear problem. We shall also refer to it as to the linear system
related to the corresponding integrable differential equation. In this 
section
we  deal with the defining representation of $sl(2)$
\a
&&H=\left(\begin{array}{cc} 1&0\\0&-1\end{array}\right),\quad
E^+= \left(\begin{array}{cc} 0&1\\0&0\end{array}\right),\quad
E^-=\left(\begin{array}{cc} 0&0\\1&0\end{array}\right).\0
\b
Therefore, the solution of the linear system (\ref{3.3a}) $\t$ is a $2\times 2$ matrix whose components depend on the spectral parameter $\l$ in the \sG\, case (\ref{3.2b}). Since $A_{\al}$ (\ref{3.2a}),
(\ref{3.2b}) are traceless, it is clear that the determinant of $\t$ does not depend on the "world--sheet" coordinates
\a
&&\del_{\mu} {\rm det}\t=0.\label{3.3b}
\bj
In what follows we shall need the notations
\a
&&A(\t)=\frac{\t_{12}}{\t_{11}},\quad B(\t)=\frac{\t_{22}}{\t_{21}}\label{3.4}
\b
where $\t_{ij},\,\,\, i,j=1,2$ are the entries of the matrix $\t$. The dependence of the quantities $A$ and $B$ on $\t$ will be skipped whenever there is no risk of confusion.

Let us first consider the \Lio\, Lax connection
\a
&&A_z=\frac{1}{2}\left(\begin{array}{cc} \frac{\del\var}{2}&e^{\frac{\var}{2}}\\0&-\frac{\del \var}{2}\end{array}\right),
\quad\quad 
A_{\bar z}=\frac{1}{2}\left(\begin{array}{cc} -\frac{\bar{\del}\var}{2}&0\\e^{\frac{\var}{2}}&\frac{\bar{\del}\var}{2}
\end{array}\right).\label{3.5}\b 
Inserting  it into (\ref{3.3a}) and taking into account
the notations (\ref{3.4}), we get the system
\ai
&&\begin{array}{ccc}\del A&=&-\frac{e^{\frac{\var}{2}}{\rm det}\t}{2\t_{11}^2}\\
\bar{\del}A&=&0\end{array} \quad\quad
\begin{array}{ccc}\del B&=&0\0\\
 \bar{\del} B&=&\frac{e^{\frac{\var}{2}}{\rm det}\t}{2 \t_{21}^2}\end{array}\\
\label{3.6a}
\b
From the above equations it is seen that the \Lio\, field is expressed as
follows 
\a
&&e^{\var (z,\bar z)}=-4 \frac{ \del A(z)\, \bar{\del} B(\bar z)}
{\left( A(z)- B(\bar z )\right)^2}
\label{3.6b}
\bj 
which resembles the \Lio\, formula (\ref{2.11}).

To treat the \sG\, equation, we  recall that the underlying connection
(\ref{3.2b}) in the defining representation of $sl(2)$ is given by the matrices
\a
&&A_+=\frac{1}{2}\left(\begin{array}{cc} i\frac{\del_+\psi}{2}&\l e^{i\frac{\psi}{2}}\\
\l e^{-i\frac{\psi}{2}}&-i\frac{\del_+ \psi}{2}\end{array}\right),
\quad\quad 
A_-=\frac{1}{2}\left(\begin{array}{cc} -i\frac{\del_-\psi}{2}&\l^{-1}e^{-i\frac{\psi}{2}}\\
\l^{-1}e^{i\frac{\psi}{2}}&i\frac{\del_-\psi}{2}
\end{array}\right)\label{3.7}\b 
 In view of (\ref{3.3a}) and 
(\ref{3.7}), we conclude that the quantities (\ref{3.4}) satisfy the equations
\ai
&&\del_{\pm}A=-\l^{\pm 1}\frac{e^{\pm i \frac{\psi}{2}} {\rm det}\t}
{2 \t_{11}^2}, \quad 
\del_{\pm}B=\l^{\pm 1}\frac{e^{\mp i \frac{\psi}{2}} {\rm det}\t}
{2 \t_{21}^2}.
\label{3.8a}
\b
In the above equations the dependence on the "world--sheet" coordinates $x^{\pm}$ and on the spectral parameter $\l$ was skipped. Using (\ref{3.8a})
it is easy to reconstruct the \sG\, field 
\a
&&e^{\pm i \psi}=-4\frac{\del_{\pm}A\del_{\mp}B}
{\left( A- B\right)^2}\0\\
&&\del_{\pm}A \del_{\pm}B=-\frac{\l^{\pm 2}}{4}\left(A-B\right)^2
\label{3.8b}
\bj
The above expressions seem to be a generalization of  the geometrical solution (\ref{2.13}).

Comparing (\ref{3.6b}) with (\ref{3.8b}), we see that there is an uniform expression for the general solution of the \Lio\, and the \sG\, equations.
In particular, both the equations are solved in terms of the functions 
$A$ and $B$ (\ref{3.4}). However, the latter are restricted by different 
conditions. In the \Lio\, case $A$ is a holomorphic function on $z$, while 
$B$ is antiholomorphic. When the \sG\, model is considered, these 
 conditions should be changed by (\ref{3.8b}).
 Note also that starting from
the equations (\ref{3.6a}) and (\ref{3.8a}) and taking into account 
(\ref{3.4}) as well as the algebraic relation
\a
&&A-B=-\frac{ {\rm det} \t}{\t_{11}\t_{21}}.\label{3.9}
\b
it turns out that the $2\times 2$  matrix $\t$ is a solution of the corresponding linear problem.
We postpone the proof of this statement to the next section where it will
shown that there is Lie--B\"acklund transformation which maps the solutions 
of the \Lio\, equation to solutions of the \sG\, equation and vice versa.
It is easy to check that (\ref{3.8b}) are sufficient to show that $\psi$ is a 
solution of the \sG\, equation. To prove this, one first observes that the identities
\a
&&\del_+\del_-A=2\frac{\del_+ A \del_- A}{A-B},\quad
\del_+\del_-B=-2\frac{\del_+ B\del_- B}{A-B}\label{3.10}
\b
follow from (\ref{3.8b}). We stress that the above identities are analogous to (\ref{2.14a}). However, in deriving (\ref{3.10}) we have only  used the zero curvature representation. The underlying \sG\, solution depends 
on additional variable $\l$. It should not be mixed with the spectral 
parameter which appears in the connection (\ref{3.2b}), (\ref{3.7}). In fact,
(\ref{3.8b}) are {\it not sufficient} to prove that $\psi$ does not depend on 
$\l$.

We proceed by discussing the symmetries of the equations (\ref{3.6a}) and
(\ref{3.8a}). It is clear that left translations $\t \rightarrow \t^g=g\t$ acting on the solutions of (\ref{3.3a}) induce gauge transformations 
$A_{\mu}\rightarrow A_{\mu}^g=-\del_{\mu} g g^{-1}+gA_{\mu}g^{-1}$.
 The functions $A$ and $B$ (\ref{3.4}) remain invariant under
left shifts by diagonal elements $g\in SL(2)$. On the other hand, it is obvious that a right
multiplication $\t\rightarrow \t g$ by an element $g$ which does not depend on the 
"world--sheet" variables where $g\in GL(2)$ for the \Lio\, model and   $g$ is in corresponding loop group $\tilde{GL(2)}$ for the \sG\,
case, leaves the linear system ({\ref{3.3a})
invariant. Setting 
\a
&&g=\left(\begin{array}{cc}\de&\be\\
\ga&\al\end{array}\right), \quad\quad \al\de-\be\ga\neq 0,\0
\b 
it is seen that right shifts induce M\"obius transformations
\a
&&A\rightarrow \frac{\al A+\be}{\ga A +\de}\quad
B\rightarrow \frac{\al B+\be}{\ga B +\de}\0\\
&&\hskip1.5cm {\rm det}\t\rightarrow {\rm det}\t\,{\rm det}g
\label{3.11}
\b
which obviously preserve the equations (\ref{3.6a}) and (\ref{3.8a}).

Up to now we have not imposed the condition of reality on the fields 
$\var$ and $\psi$. To do that we first observe that the Lie algebra $sl(2)$ 
has an involutive automorphism
\ai
&&\Pi H=-H,\quad \Pi E^{\pm}=E^{\mp}\label{3.12a}
\b
which in the defining representation is implemented by the element $\s$
\a
&&\Pi X=\s\, X\, \s , \quad 
\s=\left(\begin{array}{cc}0&1\\1&0\end{array}\right),\quad
 X\in sl(2)\label{3.12b}
\bj
Therefore, from (\ref{3.2a}) and (\ref{3.5}) it follows that
the \Lio\, field $\var$ is real iff the  following equations are satisfied
\ai
\bar{A_z}=\Pi A_{\bar z},\quad 
\bar{A_{\bar z}}=\Pi A_z\label{3.13a}
\b
where the bar stands for the complex conjugation.
In the above identities we  skipped the dependence of
$A_z$ and $A_{\bar z}$ on the "world--sheet" coordinates $z$
and $\bar z$; the generators of the $sl(2)$ algebra are assumed to be real: $\bar{H}=H,\,\,\, \bar{E^{\pm}}=E^{\pm}$.
In view of (\ref{3.12b}) and (\ref{3.13a}), one obtains the following complex conjugation rules in the defining representation
\a
\bar{A_z}=\s A_{\bar z}\s , \quad 
\bar{A_{\bar z}}=\s A_z\s\label{3.13b}
\bj
Similar involution holds for the \sG\, connection (\ref{3.2b}) for real values of $\psi$
\ai
&&\bar{A_{\pm}}(\l)=\Pi A_{\pm}(\l),\quad \l\in\RR.
\label{3.14a}
\b
In the defining representation one gets
\a
\bar{A_{\pm}}(\l)=\s A_{\pm}(\l)\s
\label{3.14b}
\bj
for real values of $\l$.
Taking into account (\ref{3.13a}) and the above equation,
we observe that whenever the matrix $\t$ satisfies 
(\ref{3.3a}) with $A_{\mu}$ given by (\ref{3.5}) or (\ref{3.7}), the element $\bar \t \, \s$ is always  a solution of the same linear system. In view of this observation, we get the complex conjugation rules
\a
&&\bar{\t}=\s \,\t\, C,\quad C\, \bar{C}=1
\label{3.15}
\b
where $C$ is independent on the "world--sheet" coordinates.
It is clear that the element $C$ is uniquely fixed by the initial data imposed on $\t$. For example, let us first suppose that at certain point $P$ of the "world--sheet"
$\t (P)=1$. Therefore, from (\ref{3.15}) it follows that
$C=\s$ and hence
\ai
&&\bar{A}=\frac{1}{B}\label{3.16a}
\b
Due to (\ref{3.3b}) and the  initial  condition imposed on $\t$, one 
concludes that ${\rm det}\t=1$. Moreover, taking into account
 (\ref{3.9}) and the above equation, 
we see that $A$ belongs to the unit disk $\DD=\{ A\in \CC;
A\bar A<1\}$. Note also that inserting back (\ref{3.16a})
into the general solution of the \Lio\, equation (\ref{3.6b}), one recovers the Poincar\'e metric on $\DD$ \cite{Dub}
\a
&&d\,s^2= e^{\var(z,\bar z)}dz\, d\bar{z}
=4\frac{dA\,d\bar{A}}{(1-A\bar{A})^2},\,\,\, |A|^2<1
\label{3.16b}
\bj

Another possible choice of initial conditions is 
$\hat{\t}(P)=\Ga$ where
\ai
&&\Ga=\frac{1}{2i}\left(\begin{array}{cc}-1&-i\\
1&-i\end{array}\right)\label{3.17a}
\b 
Taking into account  (\ref{3.12b}),  one easily verifies that the matrix $\Ga$ satisfies the commutation relation
\a
&&\Ga^{-1}\, \s\, \bar{\Ga}=1\label{3.17b}
\bj
and hence  
\ai
&&\bar{\hat{\t}}=\s\hat{\t}.\label{3.18a}
\b
Therefore, the quantities 
$\hat A$ and $\hat B$ are complex conjugated each to the 
other
\a
\bar{\hat{A}}=\hat{B}\label{3.18b}
\bj
Combining the above identity with (\ref{3.11}), (\ref{3.16a})
and setting $u=\hat{A}$, $\bar u=\hat{B}$ we get the relation \a
&&u=-i\frac{A+1}{A-1}\,, \quad |A|^2<1
\label{3.19}
\b
which provides an analytic isomorphism between the unit disk $\DD$ and the upper half plane $\HH$ \cite{La}. In particular, from (\ref{3.6b}) and (\ref{3.8b})
it follows that the expressions (\ref{2.11}) and (\ref{2.13})
provide general (local) solutions of the \Lio\, and the \sG\, equations respectively.

\renewcommand{\theequation}{\thesection.\arabic{equation}}
\setcounter{equation}{0}
\setcounter{subsection}{0}
\setcounter{footnote}{0} 
\section{Derivation of the Lie--B\"acklund Transformation}

Transformations which involve local coordinates, fields and their derivatives has been extensively studied in the literature \cite{An,Ab,Ten} in relation to the 
Lie's approach to differential equations. As  simplest example, one can quote 
the Lie tangent transformations of finite order. Here we follow the 
definitions adopted in \cite{An}. Under the assumption of invertibility, 
a classical result due to B\"acklund states that any $k^{\rm th}$ order 
tangent transformation is a prolongation of a Lie (first) order tangent 
transformation. Therefore, the Lie tangent transformations are only useful in 
the analysis of {\it first order \pdes}. There are two alternative, but 
related  each to other, approaches to study transformations between differential
equations of order higher than one. The first relies on the theory of the 
group of Lie--B\"acklund transformations which are infinite dimensional  generalization (derivatives of arbitrary order are included) of the Lie tangent transformations. On the other hand, it is possible to consider many--valued
transformations. The Bianchi--Lie transformation and its generalization 
due to B\"acklund and Darboux \cite{Ei,Bi,An,Ten} is a particular example
of such many--valued (surface) transformation. The map considered by B\"acklund has a nice
geometrical interpretation: it transforms a given surface $\S$ in $\RR^3$ into another
surface $\S^{\prime}$ in $\RR^3$.  A remarkable property of the above mentioned transformation
$\S\rightarrow \S^{\prime}$ is that to ensure the integrability,
both the surfaces $\S$ and $\S^{\prime}$ has to have the same constant negative scalar curvature.  This procedure enables, starting from a given solution 
of a fixed \pde\,, (which in this particular example is the \sG\, equation) to construct a family of solutions of the same \pde . 
The generalization of the previous geometrical construction yields to 
the general notion of B\"acklund transformation. 

The aim of the present section is to construct a Lie--B\"acklund transformation
which relates the \Lio\, equation to the \sG\, one. To introduce the notion of a Lie--B\"acklund transformation in this special case, we consider two infinite
sets of variables ${\bf L}= \{z,\,\bar z, \,\var ,\, \del \var ,\,\bar \del\var, \ldots\}$
and ${\bf S}=\{x^+,\,x^-,\,\psi , \,\del_+\psi ,\, \del_-\psi ,\, \ldots\}$ (the dots mean higher order derivatives of arbitrary order). ${\bf L}$ and   ${\bf S}$ are related to the \Lio\,
and the \sG\, equations respectively. Then according to \cite{An} a Lie--B\"acklund transformatoin is an invertible map ${\bf L}\leftrightarrow 
{\bf S}$ which preserves the tangency condition of arbitrary order and such that $\psi$ ($\var$) satisfies the \sG\, (\Lio\,) equation if and only if $\var$ ($\psi$)
is a solution of the \Lio\, (\sG\,) equation. 

We start by introducing some notations. First, let  $\t (z,\bar z)$ and 
$T(x^+, x^-,\l)$ be special solutions of the \Lio\, and the \sG\, linear 
systems (\ref{3.3a}) respectively. The components of the corresponding Lax connections
are given by (\ref{3.5}) and (\ref{3.7}). It is assumed that both the \Lio\, and the \sG\, fields 
are real. The $2\times 2$ matrices $\t$ and $T$ are  fixed by imposing the initial condition
\a
&&\t(0,0)=T(0,0,\l)=\Gamma\label{4.1}
\b
where the  matrix $\Gamma$ is given by (\ref{3.17a}). From (\ref{3.18b}) 
it follows that the quantities $A$ and $B$ (\ref{3.4}) are complex conjugated each to other.
Moreover, it has been shown in the previous section that (\ref{4.1}) implies 
that $u(\t)=A(\t)$ ($\bar{u}(\t)=B(\t)$ ) as well as $u(T)=A(T)$ ($\bar{u}(T)=B(T)$) belong to the upper half plane $\HH$. We shall further suppose that $\var$ and $\psi$ are such that 
\ai
&&u(\t)=u(T)
\label{4.2a}
\b
The above restriction can be removed by the weaker requirement that $u(\t)$ and $u(T)$ are related trough a $PSL(2,\RR)$ (or M\"obius) transformation
\a
&&u(\t)=\frac{\al u(T)+\be}{\ga u(T)+\de},\quad \al\de-\be\ga=1
\label{4.2b}
\bj
The reason for this freedom is that the solutions $\t$ and $T$ (without fixing the initial conditions (\ref{4.1})) are determined up to  right multiplication 
by an element of the group $SL(2,\RR)$. In view of (\ref{3.11}), it acts by 
M\"obius transformations on the variables $u$ and $\bar u$. As it was commented previously, the geometric interpretation of 
the ambiguity (\ref{4.2b}) is based on the fact that an isometric immersion 
is determined up to an isometry of the "target space", which in our case is the Lobachevskian plane $\HH$.  Due to  the invariance of (\ref{3.6a}) and (\ref{3.8a})
under M\"obius transformations (\ref{3.11}), we can restrict our attention
on (\ref{4.2a}) only.

Comparing (\ref{3.4}) and (\ref{3.18b}) with (\ref{4.2a}) together with 
the identities ${\rm det}\t=$\\$={\rm det}T=-\frac{i}{2}$ which are consequence 
from the initial conditions (\ref{4.1}) imposed on $\t$ and $T$, one obtains the relations
\ai
&&\t_{11}\t_{21}=t_{11}t_{21}=\frac{i}{2(u-\bar u)}
\label{4.3a}
\b
which are compatible with the identities $\bar\t_{1i}=\t_{2i};\quad\bar{t}_{1i}=
t_{2i},\quad i=1,2$. These identities follow from (\ref{3.17a})--(\ref{3.18b}).  As an output from the above relations we also deduce that the ratios $\frac{\t_{11}}{t_{11}}$ and $\frac{\t_{21}}{t_{21}}$ are 
pure phases being complex conjugated each to other
\a
&&e^{i\o}=\frac{\t_{11}}{t_{11}}, \quad
e^{-i\o}=\frac{\t_{21}}{t_{21}},\quad \o\in\RR
\label{4.3b}
\bj

Inserting back (\ref{4.2a}) into (\ref{3.6a}) and (\ref{3.8a}) we obtain
\a
&&\frac{\D (z,\bar z)}{\D (x^+,x^-)}=e^{-\frac{\var}{2}}
\left(\begin{array}{cc} \l e^{i\frac{\psi}{2}+2i\o}
&\l^{-1} e^{-i\frac{\psi}{2}+2i\o}\\ & \\
\l e^{-i\frac{\psi}{2}-2i\o}
& \l^{-1} e^{i\frac{\psi}{2}-2i\o}\end{array}\right)\0\\
&& \0\\
&&\frac{\D (x^+,x^-)}{\D (z,\bar z)}=\frac{e^{\frac{\var}{2}}}{2i\sin \psi}
\left(\begin{array}{cc} \l^{-1} e^{i\frac{\psi}{2}-2i\o}
&-\l^{-1} e^{-i\frac{\psi}{2}+2i\o}\\ & \\
\l e^{-i\frac{\psi}{2}-2i\o}
& \l e^{i\frac{\psi}{2}+2i\o}\end{array}\right)
\label{4.4}
\b
where the quantity $\o$ has been introduced trough (\ref{4.3b}) and
we have used the classical notion of Jacobian matrix: consider, say a $C^{\infty}$ map 
$y^i=y^i(x^1,x^2)$, $i=1,2$. Then the Jacobian matrix is defined by the expression
\a
&&\frac{\D (y^1,y^2)}{\D (x^1,x^2)}=\left(\begin{array}{cc}
\frac{\del y^1}{\del x^1}& \frac{\del y^1}{\del x^2}\\
\frac{\del y^2}{\del x^1}& \frac{\del y^2}{\del x^2}\end{array}\right)\0
\b
It obviously obeys the relations $\frac{\D (z^1,z^2)}{\D (y^1,y^2)}
\frac{\D (y^1,y^2)}{\D (x^1,x^2)}=\frac{\D (z^1,z^2)}{\D (x^1,x^2)}$.  The change $(x^1,x^2)\rightarrow (y^1,y^2)$ is locally 
invertible iff the associated Jacobian $J={\rm det} \frac{\D (y^1,y^2)}{\D (x^1,x^2)}$ is not vanishing.  We shall suppose that 
the matrices (\ref{4.4}) are not degenerated. Since 
\a
&&J={\rm det}\frac{\D (z,\bar z)}{\D (x^+,x^-)}=2i e^{-\var}\sin \psi
\label{4.5}
\b
we will assume hereafter that $\psi\neq 0 ({\rm mod}\pi)$.
 It is worthwhile to discuss
the geometrical meaning of the transformation $(z, \bar z)\leftrightarrow
(x^+,x^-)$. A straightforward computation based on (\ref{4.4}) tells us that
$(z,\bar z)$ are local conformal coordinates on the surface $\S$ (\ref{2.7a})
if and only if $(x^+,x^-)$ are Tchebyscheff--like coordinates (\ref{2.8a})
on the same surface.  It has been shown in \cite{Dub} that the complex coordinates $z$ and $\bar z$ considered as functions of $x^{\pm}$ satisfy the Laplace--Beltrami equation associated to the \Tsc\, metric. Let us sketch the proof of this statement. First of all we realize that 
the phase factors (\ref{4.3b}) can be eliminated. In particular, starting from (\ref{4.4}), one gets 
\ai
&&\del_+z=\l^2 e^{i\psi}\del_-z
\label{4.6a}
\b
which with the help of the identities 
\a
&&1\mp i {\rm cotg} \psi=\mp i\frac{e^{\pm i\psi}}{\sin \psi}\0
\b
can be rewritten alternatively as 
\a
&&\del_{\pm}z =\mp i \left( {\rm cotg} \psi\del_{\pm}-\frac{\l^{\pm 2}}{\sin \psi}\del_{\mp}\right) z
\label{4.6b}
\bj
The integrability of this system yields the equations
\a
&&\LL z=\LL \bar z=0\0\\
&&\LL= \l^{-2}\del_+\frac{1}{\sin \psi}\del_+ +
\l^2\del_-\frac{1}{\sin \psi}\del_- -\del_+ {\rm cotg} \psi\del_--
\del_-+{\rm cotg} \psi\del_-
\label{4.7}
\b
The operator $\LL$ is proportional to the Laplace--Beltrami operator 
$\Delta$ associated to the generalized Tchebyscheff metric (\ref{2.8a}): $\Delta=
-\frac{1}{\sin \psi}\LL$. Therefore, $z$ and $\bar z$ are zero modes of 
$\Delta$. In particular, imposing the condition that the scalar curvature
of $\S$ is  $R=-2$, it turns out that $\var$ is a solution of the \Lio\, equation and $\psi$ satisfies the \sG\, equation. However, within the differential geometry, the relation between these two equations is quite implicit. 
The reason is that to get conformal coordinates on $\S$ starting from the
Tchebyscheff ones, one has to solve   (\ref{4.7})
which is a partial differential equation of second order. On the other hand, it is possible to work with the Jacobian matrices (\ref{4.4}) in order to obtain a
Lie--B\"acklund mapping between the \Lio\, and the \sG\, models. To do that we first introduce the vectors
\a
&&v=\left(\begin{array}{c}\t_{11}\\ \t_{21}\end{array}\right),\quad\quad
w=\left(\begin{array}{c}t_{11}\\ t_{21}\end{array}\right)\label{4.8}
\b
whose components are restricted by (\ref{4.3a}) and (\ref{4.3b}).
Our first statement is the following: 

{\it Suppose that $v$ is a solution of the linear system}
\ai
&&\del v+A_z v=0 \quad\quad \bar{\del}v+A_{\bar z}v=0\label{4.9a}
\b
{\it where $A_z$ and $A_{\bar z}$ has been introduced by (\ref{3.5}). In particular, the integrability condition of the above equations is equivalent to the \Lio\, equation. Consider the change of variables $(z, \bar z)\leftrightarrow (x^+, x^-)$ defined by (\ref{4.4}). Then the vector $w$
(\ref{4.8}) is a solution of the system}
\a
&&\del_{\pm}w+ A_{\pm}w=0\label{4.9b}
\bj
{\it where $A_{\pm}$ are given by (\ref{3.7}). One then concludes that $\psi$ (\ref{4.4}) is 
a solution of the \sG\, equation}.

To prove the above assertion we first note that the identities
\ai
&&\del_+z\del_+\bar{z}=\l^2 e^{-\var}\quad\quad
\del_-z \del_-\bar{z}=\l^{-2}e^{-\var}
\label{4.10a}
\b
are consequence from (\ref{4.4}). Differentiating the first of the above equations with respect to $x^-$ and the second with respect to $x^+$, and assuming that $\del_+\del_- z=\del_-\del_+ z$ we get 
the linear algebraic system
\a
&&\left(\del_+\del_-\bar{z}, \del_+\del_- z\right)\cdot \frac{\D (z,\bar z)}{\D (x^+,x^-)}=e^{-\var }\left(\begin{array}{c} \l^2\del_- \var\\
\l^{-2}\del_+\var\end{array}\right)
\label{4.10b}
\b
which has  unique solution given by
\a
&&\del_+\del_- z=\frac{ie^{-\frac{\var}{2}-i\frac{\psi}{2}+2i\o}}{2\l\sin \psi}
\left( e^{i\psi}\del_+\var-\l^2\del_-\var\right)\0\\
&&\del_+\del_- \bar{z}=-\frac{ie^{-\frac{\var}{2}+i\frac{\psi}{2}-2i\o}}{2\l\sin \psi}
\left( e^{-i\psi}\del_+\var-\l^2\del_-\var\right)
\label{4.10c}
\bj
whenever the Jacobian (\ref{4.5}) is not vanishing. Note that the above expressions has been derived without imposing any restriction on the phase factors (\ref{4.3b}), or equivalently, on the vectors (\ref{4.8}).

The derivatives $\del_{\pm}\del_{\mp}z$ and $\del_{\pm}\del_{\mp}\bar{z}$ can
be calculated alternatively by using the Jacobian matrices (\ref{4.4}), the linear system (\ref{4.9a}) and the algebraic relations (\ref{4.3a}) and 
(\ref{4.3b}). To do that we first observe that the  expressions
\a
&&\del_{\pm}\ln \t_{11}=\pm\frac{i}{4}\left({\rm cotg}\psi\del_{\pm}\var-
\frac{\l^{\pm 2}}{\sin \psi}\del_{\mp}\var\right)-\frac{\l^{\pm 1} e^{\pm i\frac{ \psi}{2}}t_{21}}{2t_{11}}
\label{4.11}
\b
take place.
In view of the identity $\t_{21}=\bar\t_{11}$, the derivatives 
$\del_{\pm}\ln \t_{21}$ are obtained from the above equations by complex conjugation. Taking into account (\ref{4.11}) and derivating the entries of the Jacobian matrix
$\frac{\D (z,\bar z)}{\D (x^+,x^-)}$ (\ref{4.4}) with respect to $x^{\pm}$ we get the equations
\a
&&\del_{\pm}\del_{\mp}z= \l^{\mp 1}e^{-\frac{\var}{2}\mp i\frac{\psi}{2}+2i\o}
\left( \pm i \frac{e^{\pm i\psi}}{2\sin \psi}\del_{\pm}\var \mp i\frac{\l^{\pm 2}}
{2\sin \psi}\del_{\mp}\var-2\frac{\D_{\pm} t_{11}}{t_{11}}\right)
\0\\
&&\del_{\pm}\del_{\mp}\bar z= \l^{\mp 1}e^{-\frac{\var}{2}\pm i\frac{\psi}{2}-2i\o}
\left( \mp i \frac{e^{\mp i\psi}}{2\sin \psi}\del_{\pm}\var \pm i\frac{\l^{\pm 2}}
{2\sin \psi}\del_{\mp}\var-2\frac{\D_{\pm} t_{21}}{t_{21}}\right)
\label{4.12}
\b
where $\D_{\pm}$  are the covariant derivatives associated the \sG\, model
(\ref{3.2b}),  (\ref{3.7}): $\D_{\pm} t_{ij}= \left( \D_{\pm} T\right)_{ij},\,\,\,
i,j=1,2$. Due to the identity $\bar \D_{\pm} t_{11}=\D_{\pm} t_{21}$ which follows from  (\ref{3.18a}) we see that the two above equations are consistent with  the complex conjugation. Comparing (\ref{4.10c}) 
with (\ref{4.12}) we conclude that $\D_+ t_{11}=\D_-t_{21}=0$. Therefore the 
vector $w$ satisfies the equations (\ref{4.9b}). This wants to say that $\psi$ defined by (\ref{4.4}) and  (\ref{4.5}) is a solution of the \sG\, equation.

The converse is also true:

{\it Suppose that the change of the local coordinates $(z, \bar z)\leftrightarrow (x^+, x^-)$ is given by the Jacobian matrices (\ref{4.4}).
Then, imposing the equations (\ref{4.9b}) on the components $t_{i1},\,\,\,
i=1,2$ of the vector $w$ (\ref{4.8}), it turns out that 
$v=\left(\begin{array}{c} \t_{11}\\ \t_{21}\end{array}\right)$ is a solution 
of the system (\ref{4.9a}). Therefore, $\var$ satisfies the \Lio\,
equation}.

Let us sketch the proof. First, as it was mentioned before, the equations
(\ref{4.10c}) are derived  directly from (\ref{4.4}) without using  (\ref{4.9a}) and (\ref{4.9b}). On the other hand, the derivatives $\del_{\pm}
\del_{\mp}z$ and their complex conjugates can be calculated from (\ref{4.4}) 
by the use of the linear system (\ref{4.9b}). As a result one recovers the 
expressions (\ref{4.11}) and their complex conjugates. Exploiting again (\ref{4.4}) we get the identities
\a
&&\hskip-2cm {\rm cotg}\psi \del_{\pm}\var -\frac{\l^{\pm 2}}{\sin \psi}\del_{\mp} \var=
\pm i \l^{\pm 1} e^{-\frac{\var}{2}}
\left( e^{\pm i \frac{\psi}{2}+2i \o}\del\var-
 e^{ \mp i \frac{\psi}{2}-2i \o}{\bar\del}\var\right)
\label{4.13}
\b
which inserted back  into (\ref{4.11}) produce the expressions
\a
&&\del_{\pm}\ln \t_{11}=-\frac{\l^{\pm 1}e^{-\frac{\var}{2}}}{4}
\left( e^{\pm i \frac{\psi}{2}+2i \o}\del\var - 
e^{ \mp i \frac{\psi}{2}-2i \o}{\bar\del}\var\right)-
\frac{\l^{\pm 1}e^{\pm i \frac{\psi}{2}}}{2}
\frac{t_{21}}{t_{11}}\0\\
&&\del_{\pm}\ln \t_{21}=\frac{\l^{\pm 1}e^{-\frac{\var}{2}}}{4}
\left( e^{\pm i \frac{\psi}{2}+2i \o}\del\var - 
e^{ \mp i \frac{\psi}{2}-2i \o}{\bar\del}\var\right)-
\frac{\l^{\pm 1}e^{\mp i \frac{\psi}{2}}}{2}
\frac{t_{11}}{t_{21}}\label{4.14}
\b
The above equations allow us to compute  
$\D_z v$ and $\D_{\bar z} v$  where $\D_z$ and $\D_{\bar z}$ stand for the 
covariant derivatives associated to the \Lio\, connection (\ref{3.2a})  
(\ref{3.5}). In view of  (\ref{4.4}), it is seen that $\D_z v= \D_{\bar z} v=0$.   Therefore the system (\ref{4.9a}) as well as the 
\Lio\, equation take place. We then conclude that the change of coordinates on $\S$ induced by (\ref{4.4}) provides a Lie--Backl\"und transformation which relates the \Lio\, and the \sG\, equations. There is a delicate problem which needs a further investigation. To state it, we recall that the Lie--B\"acklund transformations form a Lie group $\G$. In this section we have constructed 
a special element $\gamma\in \G$ which is induced by (\ref{4.4}). However,
our analysis does not give an answer to the following question: {\it are 
$\gamma$ and the identity element in the same connected component of $\G$?}
It is clear that the existence of a continuous deformation relating the 
\Lio\, to the \sG\, equation is reduced to a positive answer of this question.

Note that the observation that (\ref{4.4}) generates a Lie--B\"acklund transformation between (\ref{2.9a}) and (\ref{2.9b}) can be derived also
from the integrability condition of the system
\a
&&i\del_{\pm}\o=\pm \frac{i}{4}\del_{\pm}\psi-
\frac{1}{4}\left(\del_{\pm}z\del \var-\del_{\pm}z\bar \del\var\right)
\label{4.15}
\b
The above equations follow from the Jacobian matrix (\ref{4.4}) and  (\ref{4.10c}) which can be written in the form
\a
\del_+\del_-z=-\del_+z\del_-z\del\var \0
\b
As a result of a straightforward calculation, one deduces that the integrability
of the equations (\ref{4.15}) is equivalent to the relation
\a
&&\frac{\del_+\del_-\psi}{\sin\psi}=2e^{-\var}\del\bar \del \var
\0
\b
which according to (\ref{2.7b}) and (\ref{2.8b}) agrees with the invariance of 
the scalar curvature under the change of the local coordinates $(z, \bar z)\leftrightarrow (x^+,x^-)$.

It is interesting to note that there is an alternative way to obtain the 
Lie--B\"acklund transformation, which we constructed in this section. To fix the idea, let us start by the \Lio\, connection (\ref{3.2a}). Under the  action of an arbitrary diffeomorphism $(z,\bar z)\rightarrow (x^+,x^-)$ where $x^{\pm}$
are real variables, it transforms as a 1-form $\DD_{\pm}=\del_{\pm}+
U_{\pm}$ where $U_{\pm}=\del_{\pm}z A_z+\del_{\pm}\bar z A_{\bar z}$. The curvature is a 2--form and therefore $\FF_{+-}=\[\DD_+ , \DD_-\]$ and  $F_{z\bar z}=\[\D_z , \D_{\bar z}\]$  are related by the equation 
$F_{z\bar z}={\rm det}\frac{\D(z,\bar z)}{\D(x^+,x^-)} \FF_{+-}$.
Denote by $g$ the element $g=e^{i\o H},\,\, \, \o\in \RR$
and consider the gauge transformation $\DD_{\pm}\rightarrow \DD_{\pm}^g=
g^{-1} \DD_{\pm} g$. Then $\DD_{\pm}^g$ coincides with \sG\, connection 
(\ref{3.2b}) provided that the Jacobian matrix of the change $(z,\bar z)\rightarrow (x^+,x^-)$ is given by (\ref{4.4}) and $\o$ satisfies (\ref{4.15}).  This approach, which will 
be presented in details elsewhere \cite{GBR}, suggests that Lie--B\"acklund transformations between integrable \pdes\, are induced by a composition of a changes of the independent variables and special
gauge transformations acting on the underlying Lax connection.

\renewcommand{\theequation}{\thesection.\arabic{equation}}
\setcounter{equation}{0}
\setcounter{subsection}{0}
\setcounter{footnote}{0} 
\section{Soliton Surfaces}
 The goal of this section is to study a subclass of pseudospherical surfaces which are related to $N$--soliton solutions of the \sG\, equation. 
Usually one considers the soliton surfaces as surfaces embedded in $\RR^3$. 
In the literature, there are known few explicit examples of soliton surfaces.
Among them one can quote the pseudospheres of Beltrami and Dini \cite{Dub,Bi,Sy}. The latter are geometric realization of the static and the moving one--soliton solutions respectively. Generic $N$--soliton surfaces has been calculated in \cite{Sy} by using appropriate Bianchi--Lie transformations    \cite{Bi,An,Ten}.
In the present section, as always within this paper, {\it we shall consider the soliton surfaces as surfaces embedded in the Lobachevskian plane $\HH$ instead 
of  surfaces embedded in $\RR^3$}. According to the analysis presented in section 3, in order to get a mapping into the upper half plane, one has to construct special solutions of the underlying linear problem which obey the complex conjugation rule (\ref{3.18a}). To get matrix solutions of the linear system (\ref{3.3a})  
related to $N$--soliton solutions of the \sG\, model, we shall use an approach proposed in \cite{Da}. Its advantage is that it can be generalized to treat quasi--periodic solutions. In what follows, for the sake of brevity we shall use the notations $f(x)=f(x^+,x^-)$ and $f(0)=f(0,0)$ for any function on the coordinates $x^{\pm}$.

First of all we observe that in order to get a {\it matrix} solution of the 
linear problem (\ref{3.3a}), (\ref{3.7}), it is enough only to know a vector solution of the corresponding linear problem. To prove this statement, we first observe that the \sG\, Lax connection (\ref{3.7}) satisfies the relations
\a
&&A_{\pm}(x,-\l)=H\, A_{\pm}(x,\l)\,H
\label{5.1}
\b
Therefore, if $w(x,\l)$ is a solution of the linear problem 
$\left(\del_{\pm}+A_{\pm}(x,\l)\right)w(x,\l)=0$, it turns out that the vector
$H w(x,-\l)$ is a solution too. From this observation we conclude that 
\a
&&W(x,\l)=\left(w(x,\l),\,\,\, H\cdot w(x,-\l)\right)
\label{5.2}
\b
is a matrix solution of the linear problem (\ref{3.3a}), (\ref{3.7}) which is related to the \sG\, equation. For generic complex values of the spectral parameter $\l$, $w(x,\l)$ and 
$H\cdot w(x,-\l)$ are independent, and therefore, they can be chosen as 
{\it fundamental} solutions of the linear system (\ref{4.9b}). Following \cite{Da}, let us suppose that for certain values $\mu_1,\ldots , \mu_N$ of 
$\l$, the $2\times 2$ matrix $W$ (\ref{5.2}) is degenerated. The integer $N$ 
coincides with the number of the solitons. The degeneracy conditions, imposed on $W(x,\l)$ mean that there are constants $c_j,\,\,\, j=1=1,\ldots , N$ such that the identities
\ai
&&w(x,\mu_j)=c_j H\cdot w(x,-\mu_j)\label{5.3a}
\b
take place. In components (cf. (\ref{4.8})) one can write
\a
&&w_k(x,\mu_j)=(-)^{k-1}c_j w_k(x,-\mu_j),\0\\
&&k=1,2,\quad j=1,\ldots, N
\label{5.3b}
\bj
The above equations has unique solution provided that one sets
\a
&&w_N(x,\l)=e(x,-\l) e^{i\Psi(x)} \left(\begin{array}{c}
\prod_{j=1}^N(\l+\ep_{1j}(x))\\
\prod_{j=1}^N(\l+\ep_{2j}(x))\end{array}\right)\0\\
&&e(x,\l)=e^{\frac{1}{2}(\l x^++\frac{x^-}{\l})}
\label{5.4}
\b 
Note that inserting back the above ansatz into (\ref{5.3b}), one gets the algebraic relations
\a
&&\prod_{l=1}^N \frac{\ep_{kj}(x)+\mu_j}{\ep_{kj}(x)-\mu_j}=
(-)^{k-1} e^2(x,\mu_j),\0\\
&&k=1,2 \quad j=1,\ldots N
\label{5.5}
\b
It has been proven in \cite{Da} that $w_N(x,\l)$ is a solution of the linear system  (\ref{4.9b}) with $A_{\pm}$ given by (\ref{3.7}) provided that
\a
&&i \del_+\psi=\sum_{l=1}^N (\del_+\ep_{1l}-\del_+\ep_{2l})\0\\
&&e^{i\psi}=\prod_{l=1}^N\frac{\ep_{2l}}{\ep_{1l}}
\label{5.6}
\b
The consistency of these equations can be proven easily by using (\ref{5.5}).
Note that a similar procedure applies equally well to the $A_n$ affine Toda solitons \cite{Beto}.

We proceed by imposing reality condition on the \sG\, field $\psi$. In view of 
(\ref{3.14b}) and the ansatz (\ref{5.4}) we conclude that 
\ai
&&w(x,\l)=\s \cdot \bar{w}(x,\bar \l)
\label{5.7a}
\b
where the element $\s$ was defined by (\ref{3.12b}). Comparing the above equation with (\ref{5.3a}) and (\ref{5.4}) we conclude that the \sG\, field is real if and only if 
\a
&&\bar{\mu_j}=\mu_{\pi(j)},\quad \bar c_j= -c_{\pi(j)}\0\\
&&\bar\ep_{1j}=\ep_{2\pi^{\prime}(j)}, \quad j=1,\ldots, N
\label{5.7b}
\bj
where $\pi$ and $\pi^{\prime}$ are two (probably different) involutive permutations of the numbers $1,\ldots , N$. 

Therefore, we can write the matrix (\ref{5.2}) as follows
\a
&&W_N(x,\l)=e^{i\Psi}\left(\begin{array}{cc}
\prod_{l=1}^N(\ep_l(x)+\l)e(-\l)& \prod_{l=1}^N(\ep_l(x)-\l)e(\l)\\
\prod_{l=1}^N(\bar\ep_l(x)+\l)e(-\l)& -\prod_{l=1}^N(\bar \ep_l(x)-\l)e(\l)
\end{array}\right)\0\\
&&\ep_l(x)=\ep_{1l}(x)\label{5.8}
\b
which by construction satisfies the linear problem associated to the \sG\, equation. Starting from the above matrix, it is easy to obtain the normalized solution
\a
&&T_N(x,\l)=W_N(x,\l)W_N^{-1}(0,\l)=
\left(\begin{array}{cc} e^{i\frac{\psi_N(x)-\psi_N(0)}{4}}X_N(\l)&
e^{i\frac{\psi_N(x)+\psi_N(0)}{4}}Y_N(\l)\\
e^{-i\frac{\psi_N(x)+\psi_N(0)}{4}}\bar Y_N(\bar\l)&
e^{-i\frac{\psi_N(x)-\psi_N(0)}{4}}\bar X_N(\bar\l)\end{array}\right)\0\\
&&X_N(\l)=\frac{\prod_{l=1}^N(\l+\ep_l(x))(\l-\bar \ep_l(0))e(-\l)+(\l\leftrightarrow -\l)}
{2\prod_{l=1}^N(\l^2-\mu_l^2)}\0\\
&&Y_N(\l)\frac{\prod_{l=1}^N(\l+\ep_l(x))(\l- \ep_l(0))e(-\l)+(\l\leftrightarrow -\l)}
{2\prod_{l=1}^N(\l^2-\mu_l^2)}\
\label{5.9}
\b
of the linear problem associated to the \sG\, equation.
In view of (\ref{3.16a}) and (\ref{3.16b}), the quantity $A(T)$ (\ref{3.4})
is in the unit disk $\DD$. Using the analytic isomorphism (\ref{3.19}) between $\DD$ and the upper half plane $\HH$ we get 
\a
&&u_N(x,\l)=i \frac{X_N(x,\l)+e^{i\frac{\psi_N(0)}{4}}Y_N(x,\l)}
{X_N(x,\l)-e^{i\frac{\psi_N(0)}{4}}Y_N(x,\l)}\label{5.10}
\b 
 The above expression gives  (up to an isometry transformation of $\HH$) an isometric immersion of an $N$--soliton surface $\S_N$ in the Lobachevskian plane.

To finish, let us consider some examples. First, suppose that one deals with the vacuum solution $\psi=0$ of the \sG\, equation (\ref{2.9b}). In this case 
$X_0(\l)={\rm ch} (\frac{\l x^+}{2}+\frac{x^-}{2\l}),\quad Y_0(\l)=-{\rm sh}(\frac{\l x^+}{2}+\frac{x^-}{2\l})$ and therefore
\a
&&u_0(x,\l)=i e^{-\l x^+-\frac{x^-}{\l}}
\label{5.11}
\b
which is a geodesic line in the Lobachevskian plane. We recall that the geodesics in the space of Lobachevski are straight lines parellel to the imaginary axis or semicircles which end on the real axis. Note that all the other geodesics in $\HH$ can be obtained from (\ref{5.11}) by a suitable $PSL(2,\RR)$ transformation (\ref{2.17}). 

To get one--soliton surfaces, we first observe that, in accordance with the general expression (\ref{5.5}), 
the (one-soliton) dynamics is governed by the equations
\ai
&&\ep_(x)=-\mu\frac{1+ce^2(\mu)}{1-ce^2(\mu)},\0\\
&&\bar\mu=\mu,\quad \bar c =-c\label{5.12a}
\b
The above equations together with (\ref{5.6}) yield
\a
&&e^{-i\frac{\psi}{2}}=-\frac{\ep}{\mu}=
\frac{1+ce^2(\mu)}{1-ce^2(\mu)}\label{5.12b}
\bj
which agree with the standard expression of the one soliton solution of the \sG\, equation \cite{Ab,inv}. Substituting back the the above solution into the general formulas (\ref{5.9}) and (\ref{5.10}) we get
\a
&&\hskip-1.5cm u_1=i\frac{\l+\mu}{\l-\mu}
\frac{(\l-\mu){\rm ch}(A(\l)+A(\mu)+\ln\al)+
i\kappa (\l+\mu){\rm sh}(A(\l)-A(\mu)}
{(\l+\mu){\rm ch}(A(\l)-A(\mu)-\ln\al)-
i\kappa (\l-\mu){\rm sh}(A(\l)+A(\mu)}\0\\
&&A(\l)=\frac{\l x^+}{2}+\frac{x^-}{2\l}
\label{5.13}
\bj
where $c_1=c=i\kappa\al$, $\al=|c|$ and $\kappa=\pm 1$.
Depending on the value of $\kappa$, the solutions (\ref{5.12a}), (\ref{5.12b}) are called solitons (for 
$\kappa=1$) and antisolitons (for $\kappa=-1$). Therefore, we conclude that the surface $\S_0$ which corresponds to the \sG\, vacuum solution is mapped into a single 
geodesic line. This is not strange since the metric on $\S_0$ is degenerated everywhere. On the other hand the 
isometric immersion $\S_1\rightarrow^{\hskip -.3cm i}\,\,\HH$  (\ref{5.13})  is not degenerated except the points at which ${\rm sin}\psi$ vanishes.
\vskip 2truecm
${\bf Acknowledgements}$ It is a pleasure to thank G. M. Sotkov and J. P. Zubelli for various stimulating discussions and for their constant interest on the present work. We are also glad to thank L. A. Ferreira  for bringing our attention on the reference \cite{Bob}.
Two of us, H. B. and G. C. acknowledge financial support 
from CNPq--Brazil. R. P. was supported partially by FAPERJ--Rio de Janeiro, and during the final stage of this work by Universidade Cat\'olica de Petropilis 
(GFT) and by FAPESP- S\~ao Paulo.

\end{document}